\documentclass[showkeys,preprintnumbers,amsmath,amssymb]{revtex4}
\usepackage{graphicx}
\usepackage{graphics}
\usepackage{epsfig}
 \def\be{\begin{equation}}
 \def\ee{\end{equation}}
 \def\ba{\begin{array}}
 \def\ea{\end{array}}

 \newcommand{\intt}{{{\int_{0}}^{\infty}}}

\begin{document}

\title {Elastic contact to nearly incompressible coatings -- Stiffness enhancement and elastic pile-up}

\author{E. Barthel$^1$, A. Perriot$^1$, A. Chateauminois$^2$ and C. Fr\'etigny$^2$ }

\address
{1) Laboratoire CNRS/Saint-Gobain "Surface du Verre et Interfaces",
39, quai Lucien Lefranc, BP 135, F-93303 Aubervilliers Cedex,
France\\ 2) Laboratoire de Physico-Chimie des Polym\`eres et des
Milieux Dispers\'es, UMR CNRS 7615 Ecole de Physique et Chimie
Industrielles (ESPCI), 10 rue Vauquelin, F-75213 Paris Cedex 05,
France}

\email{etienne.barthel@saint-gobain.com}




\date{\today}


\begin{abstract}
We have recently proposed an efficient computation method for the
frictionless linear elastic axisymmetric contact of coated bodies
[A. Perriot and E. Barthel, J. Mat. Res. 19 (2004) 600]. Here we
give a brief description of the approach. We also discuss
implications of the results for the instrumented indentation data
analysis of coated materials. Emphasis is laid on incompressible or
nearly incompressible materials (Poisson ratio $\nu>0.4$): we show
that the contact stiffness rises much more steeply with contact
radius than for more compressible materials and significant elastic
pile-up is evidenced. In addition the dependence of the penetration
upon contact radius increasingly deviates from the homogeneous
reference case when the Poisson ratio increases. As a result, this
algorithm may be helpful in instrumented indentation data analysis
on soft and nearly incompressible layers.
\end{abstract}

\maketitle

\section{Introduction}
Understanding the elastic contact to a coated substrate is a
prerequisite to the analysis of more complex thin film behaviour. It
is also an interesting problem because of two difficulties:
\begin{enumerate}
\item  the coupling by the elastic field of planar parallel
interfaces, which determines the response function;
\item the mixed boundary conditions  usually involved in contact problems.
\end{enumerate}
The problem has been dealt with in numerous publications, for
example~\cite{ElSherbiney76,Yu90,Gao92,Yoffe98,Schwarzer00}. In a
recent contribution, we have proposed a numerically efficient
approach~\cite{Perriot04}. We first summarize this technique,
insisting on the structure of the elastic response function on the
one hand and on the treatment of the mixed boundary conditions on
the other hand.

We also apply the resulting algorithm to the specific case of nearly
incompressible or incompressible layer material. Compared to
compressible layer material, the stiffness increases much more
rapidly. At the same time, the dependance of the penetration upon
contact radius is also strongly affected. More generally, the
deviation of the penetration {\it vs.} contact radius relation from
the homogeneous substrate solution could limit the direct
applicability of an Oliver-Pharr approach to instrumented
nanoindentation data treatment for coated substrates.


\section{Contact to a coated substrate}
The present method is based on the elastic response function of a
coated substrate calculated by Li and Chou~\cite{Li97}. The
expression of the response function is complex but is characteristic
of coupled planar parallel interfaces. We will first highlight these
main features.
\subsection{Response functions of coupled interfaces}
\begin{figure}
\includegraphics[width=6cm]{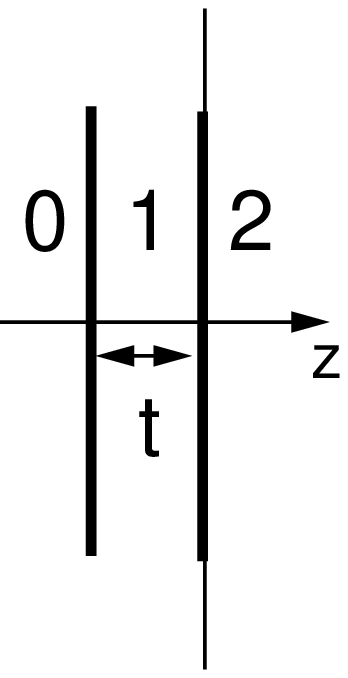}
\caption{}\label{FigSchCou}
\end{figure}

The coupling of parallel planar interfaces through the layer of
material 1 (thickness $t$) is schematized on Figure~\ref{FigSchCou}.
In-plane symmetry suggests Fourier transform in the $x,y$ plane
(wave vector $k$). If the fields obey an equation of the laplacian
type, then the response function $D$ obeys, in the notation of
Fig.~\ref{FigSchCou},
\begin{eqnarray}
\left(w_0^2-\frac{d^2}{dz^2}\right)^n D(k,z)&=&0 \mbox{   for   } z<-t \label{Lapl1}\\
\left(w_1^2-\frac{d^2}{dz^2}\right)^n D(k,z)&=&A(k)\delta(z-z') \mbox{   for   } -t<z<0 \label{Lapl2}\\
\left(w_2^2-\frac{d^2}{dz^2}\right)^n D(k,z)&=&0 \mbox{   for   }
0<z \label{Lapl3}
\end{eqnarray}
with adequate boundary conditions on the interfaces and at
$z=\pm\infty$. $A(k)$ is an excitation located on the plane $z=z'$.
For a harmonic field $n=1$, and $n=2$ for a biharmonic field. This
general structure will be specified below along with the parameters
$w_i$.

\subsubsection{Electromagnetic coupling between planar interfaces}
The interaction of the electromagnetic field radiated by polarisable
bodies at close distance results in the van der Waals interaction.
Assuming that two such media with dielectric constants $\epsilon_i$
and parallel interfaces are separated by a distance $t$, the field
fluctuations~\cite{Podgornik03} with wave vector $k$ in the $x,y$
plane and angular velocity $\omega_n$ obey
Eqs~\ref{Lapl1}-\ref{Lapl3} with
\begin{equation}
  w_i=\sqrt{\epsilon_i\omega_n^2+k^2}
\end{equation}

The response function can be calculated. For instance, the response
at $z=0$ to an excitation with wave vector $k$ located at $z=0$
(local response) is
\begin{equation}\label{VdWEq}
  D(0,0)=-\frac{2\pi}{w_1}\frac{1}{\cal D}
\end{equation}
with
\begin{equation}
  {\cal D}=1-\exp(-2w_1t)\Delta_{01}\Delta_{12}
\end{equation}
and
\begin{equation}
  \Delta_{ij}=\frac{w_i-w_j}{w_i+w_j}
\end{equation}
This form of the response function is characteristic of coupled
parallel interfaces. Indeed the denominator $\cal D$ is the
determinant resulting from the solution of the linear system
Eqs.~\ref{Lapl1}-\ref{Lapl3}. It couples the exponential field
propagation factor in medium~1 ({\em i.e.} $\exp(-2 w_1 t)$) with
the polarisability mismatch coefficients $\Delta_{ij}$. These
antisymmetric mismatch coefficients result from the boundary
conditions at the interfaces, and appear quite generally in
interfacial problems such as electrostatic image potential, Fresnel
reflection coefficients or Dundurs elastic mismatch parameters.

The response function $D$ can be used to calculate van der Waals
interactions. Indeed the secular equation ${\cal D}=0$ determines
the coupled surface plasmons eigenfrequency from which the variation
of the energy of the system with the distance $t$ can be calculated.

\subsubsection{Response function of a coated substrate}
\begin{figure}
\includegraphics[width=3.25in]{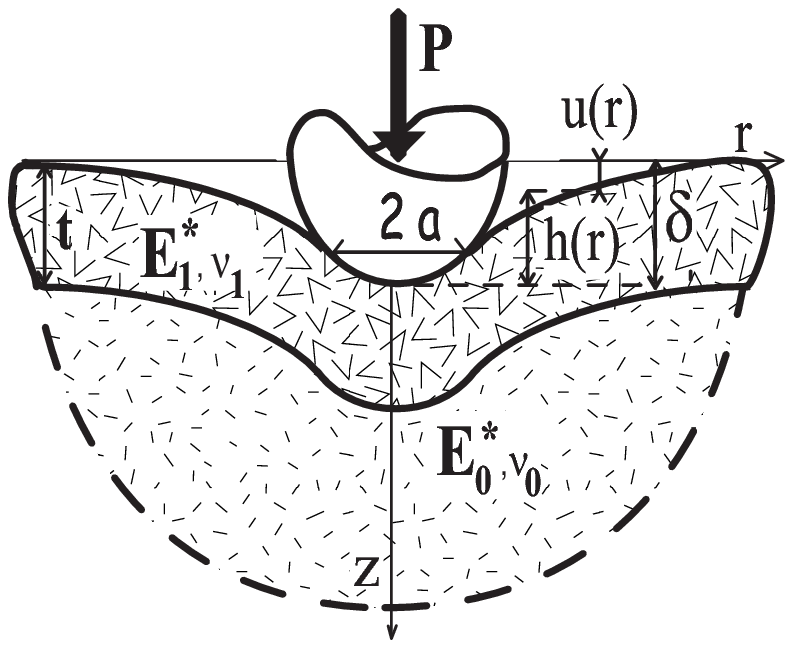}
\caption{}\label{FigSchCon}
\end{figure}

Similarly for our contact problem, we assume a coated elastic
half-space under an axisymmetric frictionless loading
(Fig.~\ref{FigSchCon}). The layer, of thickness $t$, and the
half-space are elastic, isotropic and homogeneous, while their
adhesion is supposed to be perfect. Let $E_{0}$ and $\nu_{0}$ (resp.
$E_{1}$ and $\nu_{1}$) be the elastic modulus and the Poisson ratio
of the half-space (resp. of the layer).

Due to axisymmetry, we use ${\overline{f}}$ the 0$^{th}$-order
Hankel transform of $f$ defined as\nolinebreak :
$${\overline{f}}(k)=\intt dr\,r\,J_{0}(kr)\, f(r)$$ instead of the Fourier transform.
Here $J_{0}(x)$ is the 0$^{th}$-order Bessel function of the first
kind which displays a cosine-like oscillatory behaviour suitable for
the present geometry.

Then, for a coated substrate, we fall under the framework of
Eqs.~\ref{Lapl1}-\ref{Lapl3} for linear elasticity. As is usual in
2-D elasticity problems, the fields obey a bilaplacian equation so
that $n=2$. We consider a  static problem so that $w_i=k$. Under
these hypotheses, Li and Chou obtained a relation~\cite{Li97}
between the applied normal surface stress $\overline{q}(k)$ (taken
positive when compressive) and the displacement
$\overline{u}_z(k,z)$ (positive is inwards).

More specifically for $z=0$ (with
$\overline{u}_z(k,0)\equiv\overline{u}(k)$):
\begin{equation}\label{equilHank}
k {\overline{u}}(k)=C(kt){\overline{q}}(k)
\end{equation}
where the response function is:
\begin{equation}\label{RespFonc}
C(kt)={\frac{2}{{E_1^*}}}{\frac{1+4b\,kt\,{e}^{-2kt}-ab\,{e}^{-4kt}}{1-(a+b+4b{(kt)}^{2}){e}^{-2kt}+ab\,{e}^{-4kt}}}
\end{equation}
\begin{center}
$a={\frac{\alpha\gamma_{0}-\gamma_{1}}{1+\alpha\gamma_{0}}}$,
   $b={\frac{\alpha-1}{\alpha+\gamma_{1}}}$,
   $\alpha={\frac{E_1(1+\nu_0)}{E_0(1+\nu_1)}}$,
   $\gamma_{1}=3-4{\nu}_{1}$ and $\gamma_{0}=3-4{\nu}_{0}$
\end{center}
${E_1^*}$ being the reduced modulus of the layer defined as
$E_1/(1-{\nu_1}^2)$.

The structure of the response function is again characterized by the
exponential field propagation factors and antisymmetric elastic
mismatch factors for the field transmission at the interfaces.
However, compared to electromagnetic response, the form is
complexified by the bilaplacian-type field and the shear-free
boundary conditions at the surface.

\subsection{Boundary conditions} Contact problems such as indentation
are characterized by mixed boundary conditions
(Fig.~\ref{FigSchCon}): the normal surface displacement $u$ is
specified in the contact zone -- by the penetration and shape of the
indenter -- while the normal applied stress $q$ is given outside --
usually zero --. Now Eq.~(\ref{equilHank}) can only be used to
calculate the surface displacement everywhere on the surface
provided the normal surface stress is known everywhere on the
surface. The response function Eq.~\ref{RespFonc} is therefore
useless as such~\cite{Li97}. However, this problem can be
circumvented by a mathematical trick (or change of basis, or use of
auxiliary functions).

The method, which appears explicitly in the later papers by
Sneddon~\cite{Sneddon65,Yu90}, has proved fruitful for the
description of complex contacts. For example we have been able to
propose an exact solution to the long standing question of the {\em
adhesive} contact of viscoelastic spheres in this
way~\cite{Haiat03}.

Algebraically the method relies on the relation
\begin{equation}\label{costra}
  J_0(kr) = \int_0^r\frac{\cos(kt)dt}{\sqrt{r^2-t^2}}
\end{equation}
that is to say the Bessel function $J_0$ is the cosine transform of
the function
\begin{equation}
 \frac{\Theta(r-t)}{\sqrt{r^2-t^2}}
\end{equation}
where $\Theta$, the Heaviside step function, anticipates our mixed
boundary conditions.

The trick can also be viewed as transforming back by a {\em cosine
transform} the variables which were transformed forward with a {\em
Hankel transform}. We do not fall back in proper real space, but in
a space where our boundary conditions assume a suitable form, due to
Eq.~\ref{costra}.

Explicitly we introduce the auxiliary fields $g$ and $\theta$
defined as~\cite{Haiat03}:
\begin{equation}\label{g_Hank}
g(s)=\intt dk\, {\overline{q}}(k)\cos (ks)
\end{equation}
\begin{equation}\label{th_Hank}
\theta(s)=\intt dk\, k {\overline{u}}(k)\cos (ks)
\end{equation}
Let us now apply the cosine transform to Eq. (\ref{equilHank}).
Then\nolinebreak:
\begin{equation}\label{equil_couche}
\theta(s)={\frac{2}{\pi}}\intt  g(r) \left(\intt dk\, C (kt) \cos
(kr) \cos (ks)\right)dr
\end{equation}
Similarly, Eq.~\ref{g_Hank} becomes
\begin{equation}
g(t) = \int_{t}^{+\infty} \frac{s q(s)ds}{\sqrt{s^2-t^2}}
\label{gext}
\end{equation}
As a result of the boundary condition $q(s)=0$ for $s>a$, $g(t)=0$
for $t>a$ where $a$ is the contact radius. Then the upper bound in
the spatial integral Eq.~\ref{equil_couche} is actually $a$, not
infinity. The equilibrium equation~\ref{equil_couche} has now the
form
\begin{equation}\label{equil_couche2}
\theta(s)={\frac{2}{\pi}}\int_0^a  g(r) K(r,s)dr
\end{equation}
and a simple algorithm can then be devised.

Indeed the function $\theta$ is known  inside the contact zone for
an arbitrary indenter shape $h(r)$: from Eqs.~\ref{costra} and
\ref{th_Hank} one obtains ($s<a$)
\begin{equation}\label{th_real}
\theta(s)={\frac{d}{ds}}{{\int_{0}}^{s}}dr
{\frac{rh(r)}{\sqrt{{s}^{2}-{r}^{2}}}}
\end{equation}
This can be calculated analytically for most simple indenter shapes.
In addition, the kernel $K$ in Eq.~\ref{equil_couche2}, as defined
by Eq.~\ref{equil_couche}, can be calculated from the response
function $C(kt)$ by fast Fourier transform (FFT).

Thus, discretizing Eq.~\ref{equil_couche}, $g(s)$ for $s<a$ can be
calculated numerically as the solution to a linear system. From
$g(s)$, $s<a$, force, penetration and contact stiffness can be
computed~\cite{Perriot04}.

Some of the results obtained with this algorithm have been presented
earlier. In this paper, we want to insist on a specific case: when
the layer is incompressible or nearly incompressible.

\section{Incompressible and nearly incompressible coatings - effective contact
stiffness}\label{SecEffS} In our previous paper~\cite{Perriot04}, we
calculated the contact force $P$, the contact stiffness $S$ and the
penetration $\delta$ as a function of contact radius $a$ for a large
range of modulus mismatch and a {\em single value} of Poisson ratio
$\nu=0.25$ for both substrate and film. These results provided a
detailed description of the transition between the film dominated
regime and the substrate dominated regime.

It is well known, however, that confinement of the layer at large
contact radius values $a/t\gg1$ will result in specific phenomena
for nearly incompressible materials~\cite{Ganghoffer95}. Indeed, for
such materials, volumetric deformations are penalized so that shear
deformations predominate. In the present case, such deformations are
hampered by the confinement. The response is therefore dependent
upon the axial compression modulus in the absence of lateral strain
-- the so-called \oe dometric modulus of the layer~\cite{Gacoin06}
--
\begin{equation}
  E_o=\frac{E(1-\nu)}{(1-2\nu)(1+\nu)}
\end{equation}

Especially relevant is the case of a compliant layer deposited on a
more rigid substrate. This could be a polymer film on a glass
substrate for instance. Using our previous normalization
scheme~\cite{Perriot04}, we have calculated the response of the
elastic frictionless contact of a sphere on a coated substrate for a
film/substrate reduced modulus mismatch equal to 0.1. The numerics
were carried out with the Igor data treatment software (Wavemetrics)
on a standard 1.60 GHz processor. To probe the relevant part of the
response function, a variable cut-off depth $B=3000 a/t$ was used.
The other parameters for the numerics are $N_1=2^{20}$ points for
the cosine transform and the contact radius $a$ was discretized over
$N_2=500$ points. The resulting computation time for a given contact
radius is of the order of 1 second. The results at large Poisson
ratios were checked by increasing the computation parameters $N_1$
and $N_2$ without significant variations in the results.

Fig.~\ref{FigEeq} displays the normalised contact stiffness, or
effective reduced modulus, as a function of the normalized contact
radius for Poisson ratios comprised between 0.1 and 0.5. The usual
transition between film modulus (0.1) and substrate modulus, which
is the normalizing parameter, {\it i.e.} 1, is readily obtained. The
results are almost insensitive to the Poisson ratio $\nu$ for
$\nu<0.25$. Sizeable deviation is recorded for $0.4<\nu<0.5$. In
this range, we observe that the transition occurs earlier and is
increasingly steep as the layer becomes less compressible. However,
the system does finally reach the bare substrate reduced modulus.

For incompressible coating materials, this earlier saturation of the
reduced modulus is essentially due to the elastic compliance of the
substrate itself and is better evidenced from the surface and
interface normal displacements. These displacements as computed by
inverse Hankel transform from the discrete $g$ functions obtained at
various $a/t$ ratios are displayed on Fig.~\ref{FigDeflex}. At low
penetrations ($a/t=0.1$), the substrate remains undeformed and the
compressible and incompressible systems are undistinguishable. At
larger contact radii ($a/t=1$), the substrate starts to deform and
the coating material is increasingly confined. For the
incompressible material, an {\em elastic} pile-up of the coating
material forms around the contact zone, while a global elastic
sink-in results from the deformation of the substrate and increases
with increasing contact radius. Generally speaking, these results
illustrate the impact of the incompressibility on the partitioning
of the elastic displacements between substrate and coating. Most
noteworthy, at contact radius $a/t=8$, for an {\em incompressible}
layer, the surface displacement is almost completely due to {\em
substrate} deformation; yet a roughly equal contribution from layer
and substrate is calculated for a {\em compressible} coating
material under the same conditions of confinement.

\section{Surface deflexion and instrumented indentation data analysis}
With instrumented indentation, in the absence of direct measurement
of the contact area, the Oliver \& Pharr method is used to infer the
contact radius from measured variables, {\it i.e.} force,
penetration and contact stiffness~\cite{Oliver92}. One usually uses
\begin{equation}\label{OP}
    h_c=\delta - \varepsilon {\frac{P}{S}}
\end{equation}
where $S$ is the contact stiffness, $P$ the force, $\delta$ the
penetration and $h_c$ the contact depth. The form of the equation
and the value of the constant $\varepsilon$ are such that
eq.~\ref{OP} is exact for a purely elastic system. It has been
experimentally demonstrated that eq.~\ref{OP} is very useful also
for elasto-plastic materials, mostly when {\em plastic} pile-up is
minimal.


For the analysis of instrumented indentation on coated substrates,
the same equation is used. However, Eq.~\ref{OP} is no longer valid.
Similarly, the usual relation between penetration and contact radius
such as
\begin{equation}\label{EqPen}
  \delta=\frac{\pi}{2}\frac{a}{\tan \beta}
\end{equation}
for a cone of half-included angle $\beta$ breaks down. It is
replaced by a more complex relation which involves the mechanical
parameters of the system. This explicits the fact that
Eq.~\ref{EqPen} is a {\em mechanical} relation even though the
mechanical parameters have accidentally dropped out for a
homogeneous substrate. For coated substrates, some examples are
plotted on Fig.~\ref{FigDnorm}. A salient feature is that small
contact radii are more easily reached ({\it} i.e. with smaller
penetrations) for incompressible layers than compressible layers.
This is due to the elastic pile-up effect. However, for
$a/t\simeq4.5$ a cross-over takes place. Larger radii are achieved
by larger penetrations for incompressible layers because of the
increased effective stiffness.

If we keep the form of Eq.~\ref{OP} for coated substrates, we may
attempt a more accurate approach by {\em calculating} the value of
$\varepsilon$ for a given coating configuration. Some results are
displayed on Fig.~\ref{FigEps}. Note that in the transition region,
$\varepsilon$ drops to significantly lower values and that again the
approach to incompressibility results in considerable variations.
Values for $\varepsilon$ as low as $0.2$ may affect the data
treatment in cases where the elastic contribution is significant
(large $S$, {\em ie} sizeable elastic recovery). Due to the ease of
calculation involved here, adequate $\varepsilon$ values are readily
calculated self-consistently during actual data treatment.

Finally, it is important to observe that failure to reach a correct
evaluation of the contact radius is a double source of errors. A
direct error incurs on the evaluation of the {\em effective reduced
modulus} ${E}^*$ through
\begin{equation}
    {E}^*={\frac{\sqrt{\pi}S}{2\beta \sqrt{A}}}
\end{equation}
where $\beta$ is the non-axisymmetry correction factor and $A$ the
contact area. A second error affects the evaluation of the {\em
layer reduced modulus} from the effective reduced modulus because
the model for the latter, such as discussed in section
\ref{SecEffS}, will be evaluated at an incorrect value of the
contact radius. In the case of compliant films, these errors add up.
An evaluation for $\nu=0.2$ and a modulus mismatch of 100 results in
an error of up to 60~\% at $a/t\simeq3$.

\section{Conclusion}

Summarizing our approach to the elastic contact of coated
substrates~\cite{Perriot04}, we have shown that the form of the
response function calculated by Li and Chou~\cite{Li97} is
characteristic of the coupling of parallel planar interfaces through
the elastic field. Using a new basis of functions well suited to the
mixed boundary conditions, we obtain an integral relation that is
easily solved numerically and provides force, penetration and
contact stiffness as a function of the contact radius, for arbitrary
moduli ratios and arbitrary axisymmetric indenter shape.

The stiffness of incompressible layers increases much more steeply
with contact radius than more compressible coatings as soon as the
layer is confined in the contact. Indeed volumetric deformation is
strongly penalized while significant shear is prevented by the
confined geometry.

This is accompanied by sizeable {\it elastic} pile-up which
significantly {\it reduce} the penetration needed to reach a given
contact radius. For large contact radii, the stiffness of the layer
dominates and the surface deformation is almost completely due to
the elastic yielding of the substrate. Again, this phenomenon occurs
at lower contact radii for nearly incompressible layers because of
the enhanced stiffness.

The deviation from the homogeneous material penetration {\it vs.}
contact radius relation presumably bears upon the validity of the
Oliver and Pharr approach to instrumented indentation. A possible
improvement for soft and quite elastic layers could be obtained by
calculating more accurate values of $\epsilon$ with the present
method.

\section*{ACKNOWLEDGMENTS}
The authors thank the Saint-Gobain group for its interest in
fundamental thin film problems.

\bibliographystyle{unsrt}
\bibliography{D:/data/Biblio/Capillarite/capillarity,D:/data/Biblio/MecaniqueCouches/MecaCouche,D:/data/Biblio/MecaSolGel/MecaSolGel,D:/data/Biblio/Mecanique/Mecanique,D:/data/Biblio/Materiaux/Materiaux,D:/data/Biblio/Divers/eb_gen5,D:/data/Biblio/vanderWaals/ForcesdeSurface}

\begin{thebibliography}{10}

\bibitem{Gao92}
H.~J. Gao, C.~H. Chiu, and J.~Lee.
\newblock {\em Int. J. Solids Structures}, 29, 2471, 1992.

\bibitem{Schwarzer00}
N.~Schwarzer.
\newblock {\em J. Tribology}, 122, 672, 2000.

\bibitem{Yoffe98}
E.~H. Yoffe.
\newblock {\em Phil. Mag. Let.}, 77, 69, 1998.

\bibitem{Yu90}
H.~Y. Yu, S.~C. Sanday, and B.~B. Rath.
\newblock {\em J. Mech. Phys. Sol.}, 38, 745, 1990.

\bibitem{ElSherbiney76}
M.~El-Sherbiney and J.~Halling.
\newblock {\em Wear}, 40, 325, 1976.

\bibitem{Perriot04}
A.~Perriot and E.~Barthel.
\newblock {\em J. Mat. Res.}, 19, 600, 2004.

\bibitem{Li97}
J.~Li and T.~W. Chou.
\newblock {\em Int. J. Solids Structures}, 34, 4463, 1997.

\bibitem{Podgornik03}
R.~Podgornik, P.~L. Hansen, and V.~A. Parsegian.
\newblock {\em J. Chem. Phys.}, 119, 1070, 2003.

\bibitem{Sneddon65}
I.~N. Sneddon.
\newblock {\em Int. J. Engng. Sci.}, 3, 47, 1965.

\bibitem{Haiat03}
G.~Haiat, M.~C.~Phan Huy, and E.~Barthel.
\newblock {\em J. Mech. Phys. Sol.}, 51, 69, 2003.

\bibitem{Ganghoffer95}
J.~F. Ganghoffer and A.~N. Gent.
\newblock {\em J. Adh.}, 48, 75, 1995.

\bibitem{Gacoin06}
E.~Gacoin, C.~Fretigny, A.~Chateauminois, A.~Perriot, and E.~Barthel.
\newblock to appear in {\em Trib. Let.}.

\bibitem{Oliver92}
W.~C. Oliver and G.~M. Pharr.
\newblock {\em J. Mater. Res.}, 7, 1564, 1992.

\end{thebibliography}

\newpage

\section*{CAPTIONS}

{\underline{Fig.~1:}} Geometry of the coupling between parallel planar interfaces. The layer thickness is $t$.\\

{\underline{Fig.~2:}} Schematic representation of the indentation of a coated elastic half-space.\\

{\underline{Fig.~3:}} Effective reduced modulus $S/2a$ as a function
of normalized contact radius $a/t$ for a layer/substrate reduced
modulus mismatch of 0.1 ($E^{\star}_{substrate}=1$,
$E^{\star}_{film}=0.1$) and different layer Poisson ratios. The
substrate Poisson ration is 0.2.\\

\begin{figure}
\includegraphics[width=12cm]{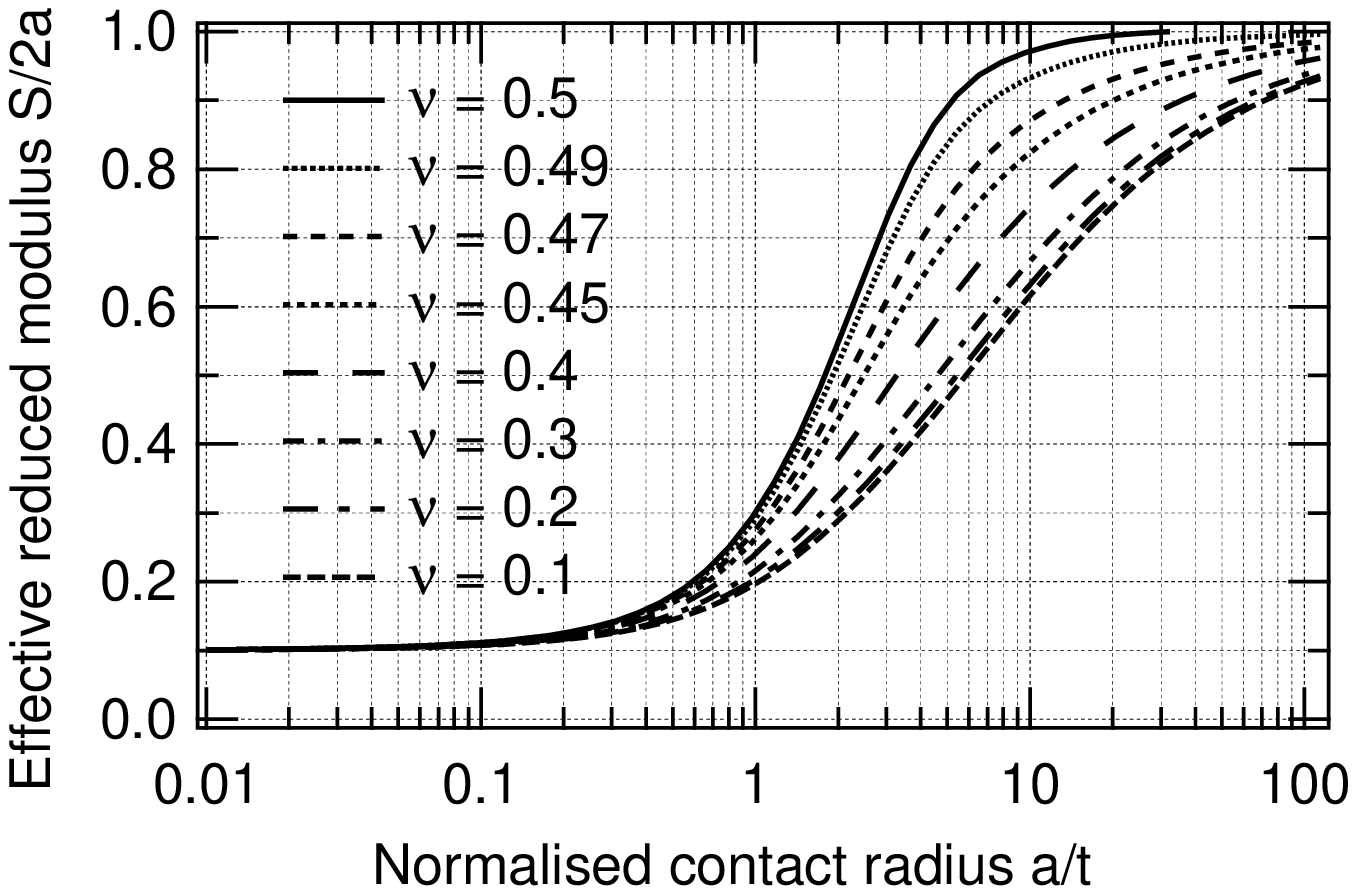}
\caption{}\label{FigEeq}
\end{figure}

\begin{figure}
\includegraphics[width=12cm]{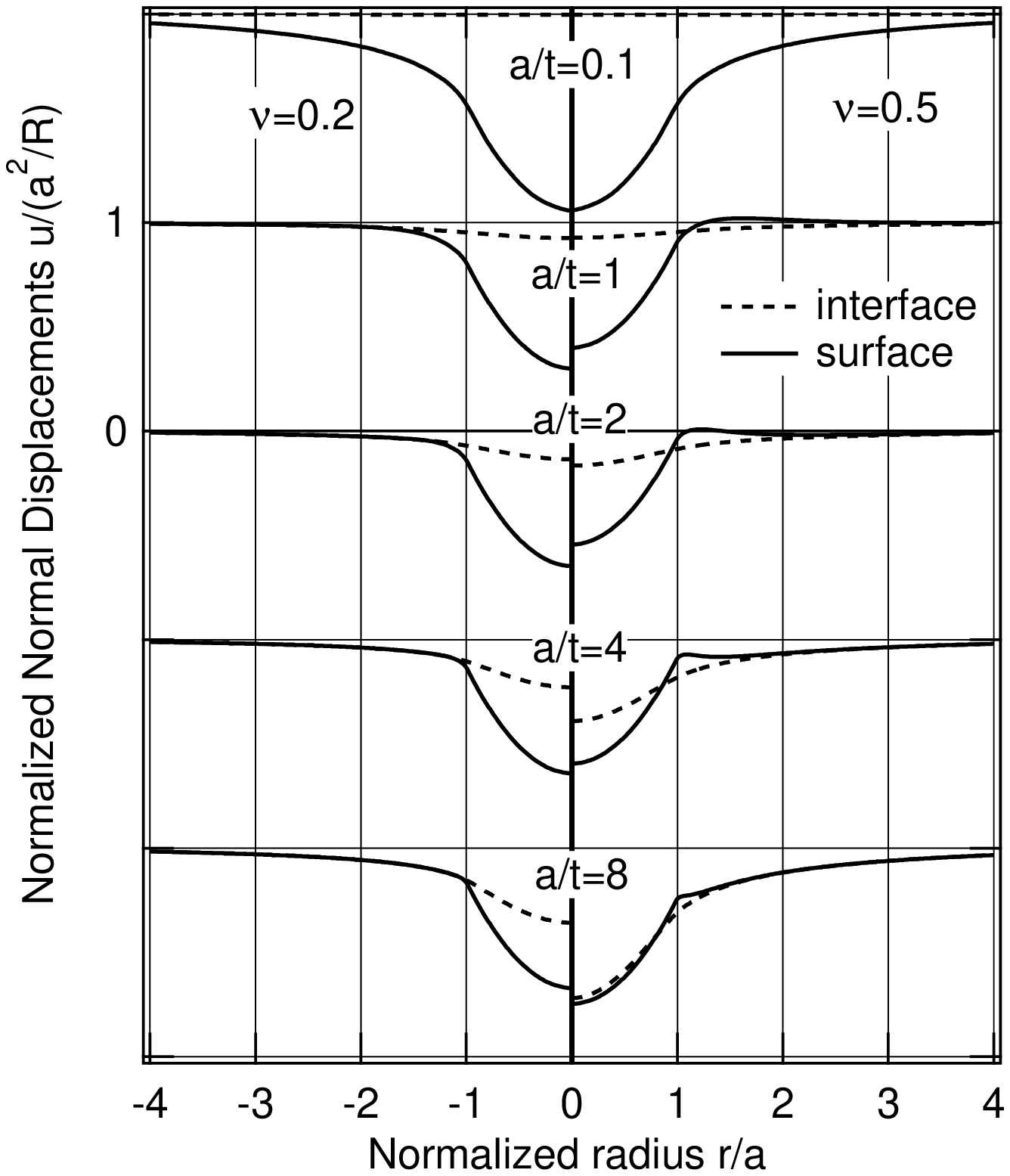}
\caption{}\label{FigDeflex}
\end{figure}

\begin{figure}
\includegraphics[width=12cm]{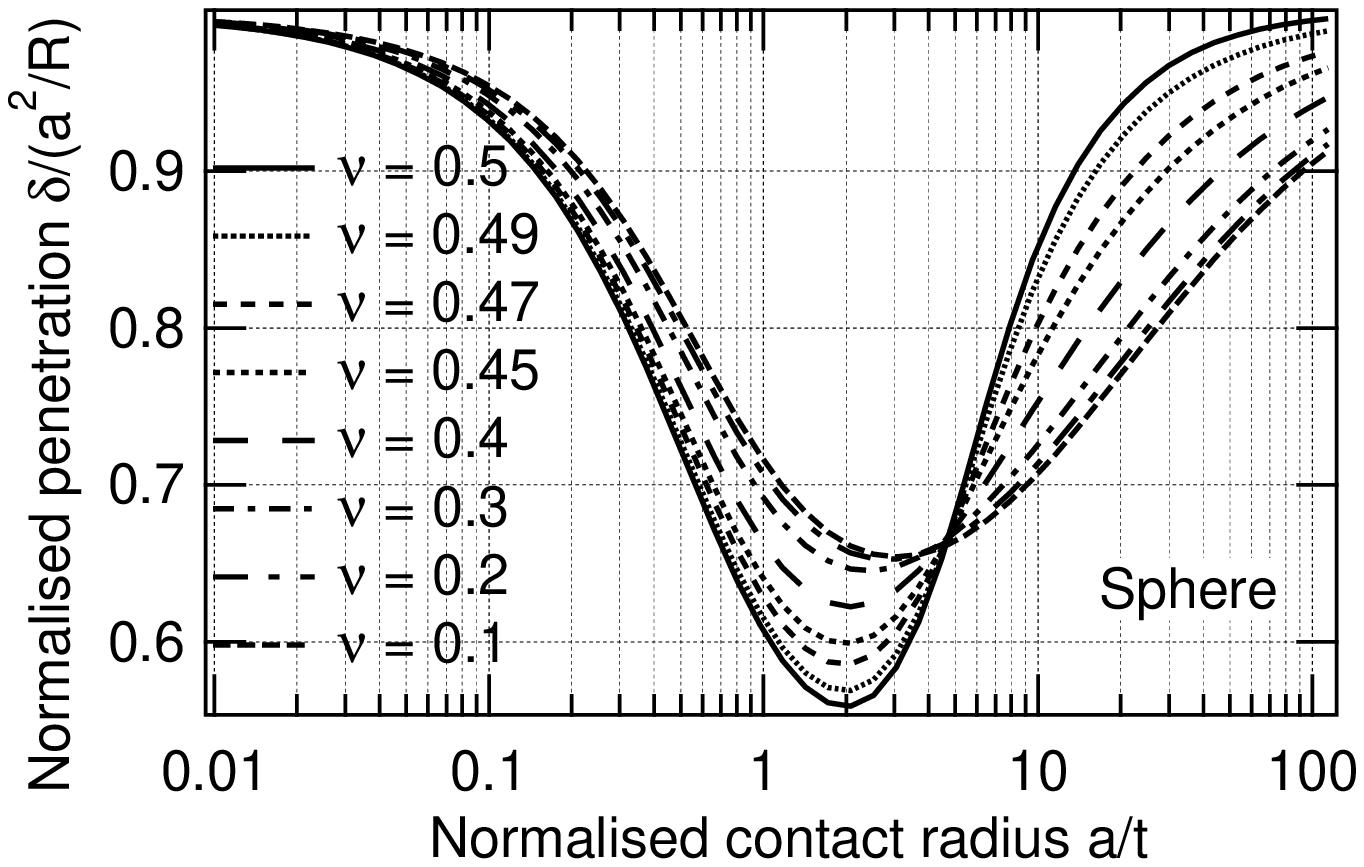}
\caption{}\label{FigDnorm}
\end{figure}

\begin{figure}
\includegraphics[width=12cm]{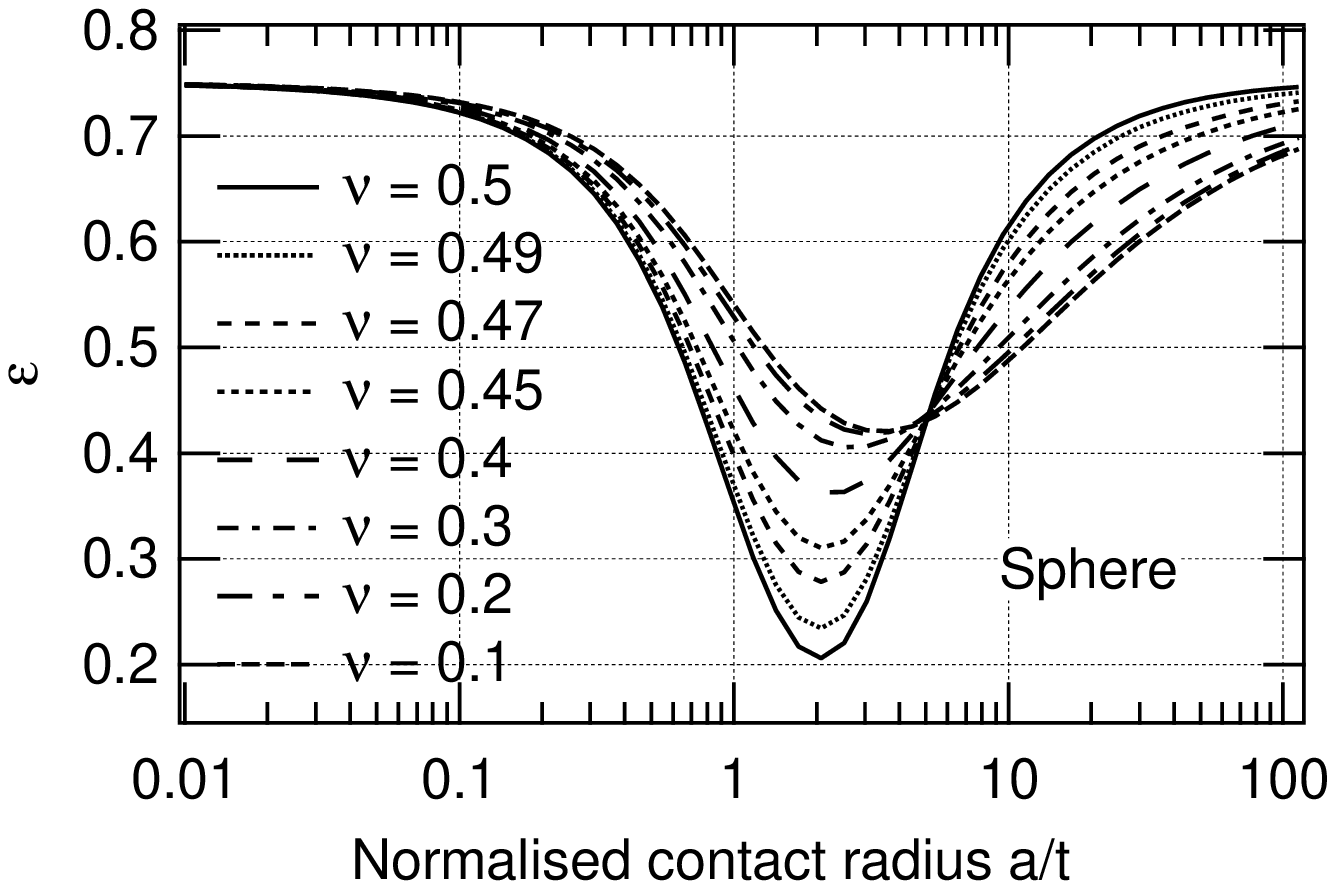}
\caption{}\label{FigEps}
\end{figure}

{\underline{Fig.~4:}} Computed normal displacement $u$ of surface
(plain) and interface (dashed) for the contact of a sphere to a
coated substrate ($E^{\star}_{substrate}/E^{\star}_{film}=10$).
Radii are normalized to contact radius $a$, displacements to
$a^2/R$. The coating Poisson ratio is $\nu=0.2$ on the left, $0.5$
on the right. Calculation parameters as in Fig.~\ref{FigEeq}.
The curves are offset by unit increments for clarity.\\

{\underline{Fig.~5:}} Computed penetration as a function of
normalized contact radius for a sphere on a coated substrate
($E^{\star}_{substrate}/E^{\star}_{film}=10$, $\nu_{substrate}=0.2$) for various film Poisson ratio. The penetration is normalized to $a^2/R$.\\

{\underline{Fig.~6:}} Computed values for the "constant" $\epsilon$
in Eq.~\ref{OP}
as a function of normalized contact radius and film Poisson ratio. Same parameters as in Fig.~5.\\

\newpage

\newpage

\end{document}